\documentclass{aastex}
\usepackage{spr-astr-addons}
\usepackage{url}

\def\beq{\begin{eqnarray}}
\def\eeq{\end{eqnarray}}

\begin{document}

\title{Revisiting vertical structure of neutrino-dominated accretion disks:
Bernoulli parameter, neutrino trapping and other distributions}

\shorttitle{Revisiting vertical structure of neutrino-dominated accretion disks}
\shortauthors{Liu et al.}

\author{Tong Liu, Wei-Min Gu, Li Xue and Ju-Fu Lu}
\affil{Department of Physics and Institute of Theoretical Physics
and Astrophysics, Xiamen University, Xiamen, Fujian 361005, China}

\email{tongliu@xmu.edu.cn}

\begin{abstract}
We revisit the vertical structure of neutrino-dominated accretion flows (NDAFs) in spherical coordinates with a new boundary condition based on the mechanical equilibrium. The solutions show that NDAF is significantly thick. The Bernoulli parameter and neutrino trapping are determined by the mass accretion rate and the viscosity parameter. According to the distribution of the Bernoulli parameter, the possible outflow may appear in the outer region of the disk. The neutrino trapping can essentially affect the neutrino radiation luminosity. The vertical structure of NDAF is like a ``sandwich'', and the multilayer accretion may account for the flares in gamma-ray bursts.
\end{abstract}


\section{Introduction}

Neutrino-dominated accretion flows (NDAFs) involve a stellar black hole of $2\sim 10 ~M_\odot$, accreting wih a hypercritical rate, in the range of $0.01 \sim 10 ~M_\odot ~{\rm s}^{-1}$. In the past decade, a number of studies have investigated this model (\emph{e.g.} \citet{Popham1999,Narayan2001,Kohri2002,Di Matteo2002,Rosswog2003,Kohri2005,Lee2005,Gu2006,Chen2007,Janiuk2007, Liu2007, Liu2008, Liu2010a, Liu2010b}). The model can provide a good understanding both the energetics of gamma-ray bursts (GRBs) and the processes of making the relativistic and baryon-poor fireballs by neutrino annihilation or magnetohydrodynamic processes (\emph{see e.g.} \citet{Popham1999} and \citet{Di Matteo2002} for references).

In cylindrical coordinates, \citet{Gu2007} discussed the underlying importance of taking the explicit form of the gravitational potential for calculating slim disk solutions, and pointed out that the H\={o}shi form of the potential \citep{Hoshi1977} is valid only for geometrically thin disks with $H/R \la 0.2$. \citet{Liu2008} found that NDAFs have both a maximal and a minimal possible mass accretion rate at their each radius, and presented a unified description of all the three known classes of optically thick accretion disks around black holes, namely Shakura-Sunyaev disks, slim disks, and NDAFs. These two works are, however, based on the simple one-zone vertical hydrostatic equilibrium. Furthermore, \citet{Gu2009} revisited the vertical structure of advection-dominated accretion disks in spherical coordinates and showed that those disks should be geometrically thick rather than being slim. As a continuation, \citet{Cai2010} found an analytic relation ${c_{\rm s}}_0/v_{\rm K} \Theta=[(\gamma-1)/2 \gamma]^{1/2}$ for the vertical mechanical equilibrium, where ${c_{\rm s}}_0$ is the sound speed on the equatorial plane, $v_{\rm K}$ is the Keplerian velocity, $\Theta$ is the half-opening angle of the flow, and $\gamma$ is the adiabatic index. However, the detailed radiation cooling was not considered in that work, and therefore no thermal equilibrium solution was established. Thus, \citet{Liu2010a} presented the vertical structure of NDAF calculating with the numerical method of \citet{Gu2009}. We found the luminosity of neutrino annihilation is enhanced by one or two orders of magnitude. The empty funnel along the rotation axis can naturally explain the neutrino annihilable ejection.

The Bernoulli parameter, defined as the summation of kinetic energy, potential energy and enthalpy of the accreting gas, is an essential quantity in accretion flows since it can be used to measure the possibility of arising outflows or winds \citep{Narayan1994,Narayan1995}. It is well-known that the positive Bernoulli parameter is necessary for the occurrence of outflow. \citet{Narayan1995} considered rotating spherical accretion flows ranging from the equatorial plane to the rotation axis. They found that the Bernoulli parameter is positive in advection-dominated conditions, especially close to the rotating axis. Here we also can evaluate the vertical distributions of Bernoulli parameter basing on our solutions, and give a judgement for whether possibly arising outflows.

Moreover, the neutrino trapping is another notable point in the studies of NDAFs. \citet{Begelman1978} presented photon trapping in Bondi accretion. \citet{Ohsuga2002,Ohsuga2003,Ohsuga2005} and \citet{Ohsuga2011} discussed the photon trapping in black hole accreting systems. For NDAFs, \citet{Di Matteo2002} argued that the neutrino trapping occurs when the neutrino optical depth is larger than $2/3$, and the inner regions of their solutions are all opaque for neutrinos even for the moderate accretion rate due to the using of Newtonian potential. Thus, in their solutions, neutrinos are significantly trapped by the flow, and the neutrino annihilation luminosity becomes too low to boost GRBs. \citet{Gu2006} found that the neutrino optical depth would significantly decrease and there would be no any neutrino trapping if the general relativistic effect is considered. Accordingly, they argued that NDAFs still have enough luminosity to work as the central engine of GRBs. \citet{Chen2007} discussed the neutrino trapping radius in NDAFs. They found that this radius depends on the accretion rate, viscosity parameter and spin of black hole. These works all concentrated on discussing neutrino trapping in the equatorial plane of disks, so it is necessary to study the neutrino trapping in the vertical structure of NDAFs.

Our purposes of this paper are to investigate the vertical structure, distributions of Bernoulli parameter and neutrino trapping in NDAFs with detailed neutrino radiation and new boundary condition. In section 2, we show the numerical results obtained by solving the vertical differential equations with the self-similar assumptions in the radial direction and a boundary condition on disk surface, then we give the definition of  Bernoulli parameter and  the criterion of neutrino trapping. In section 3, we show the vertical distributions of the Bernoulli parameter, neutrino trapping and other physical quantities, such as radial velocity and electron fraction of different cases. The possible applications in GRBs of these results is also mentioned. Conclusions and discussions are made in section 4.

\begin{figure*}
\centering
\includegraphics[angle=0,scale=0.25]{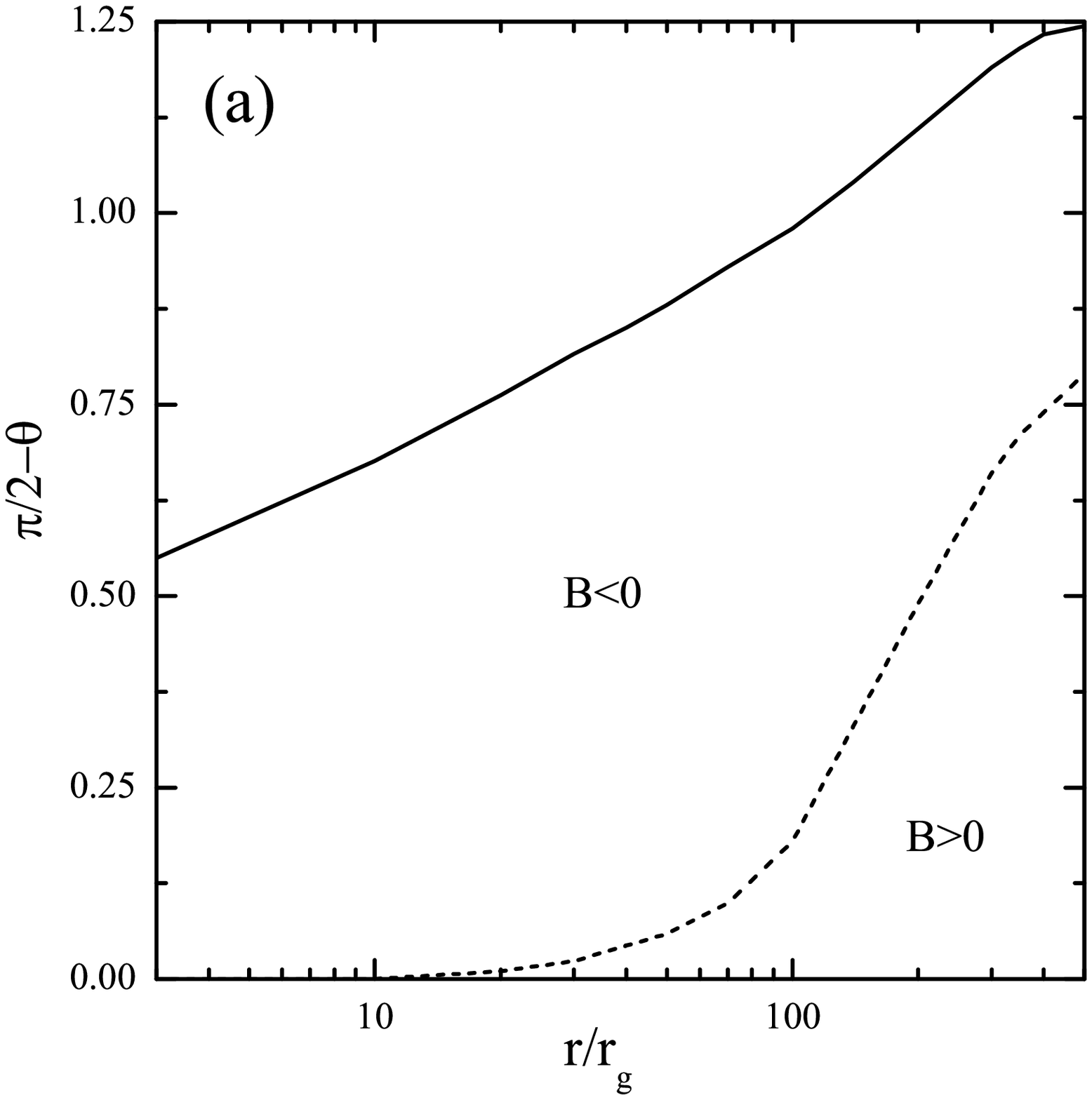}
\includegraphics[angle=0,scale=0.25]{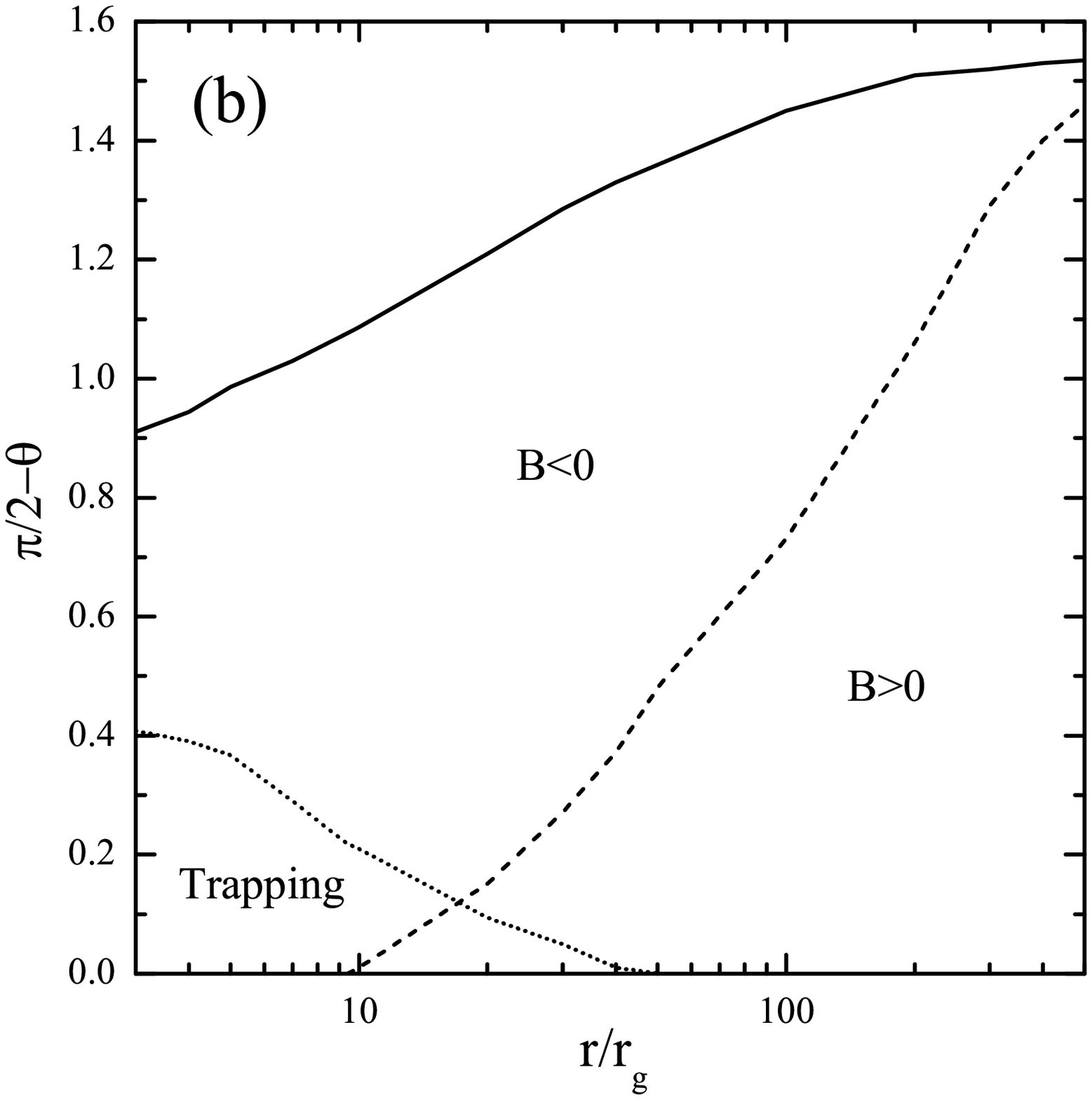}
\caption{Variations of the half-opening angles of the disk surface $(\pi/2-\theta_0)$ (solid lines), $B=0$ (dashed lines), $\tilde{t}=1$ (dotted line) with radii $r/r_g$ for which the given parameters are set as (Case 1) $\alpha=0.1$ and $\dot{M}=1~M_\odot ~\rm s^{-1}$, comparing (a); (Case 2) $\alpha=0.01$ and $\dot{M}=10~M_\odot ~\rm s^{-1}$, comparing (b).}
\label{sample-figure3}
\end{figure*}

\section{Equations and Boundary Conditions}

\subsection{Equations}

We consider a steady state axisymmetric accretion flow in spherical coordinates ($r$, $\theta$, $\phi$), i.e., $\partial/\partial t =\partial/\partial \phi = 0$. We adopt the Newtonian potential $\psi = - GM/r$, since it is convenient for self-similar assumptions, where $M$ is the mass of the central black hole. The basic equations are composed of continuity, momentum and energy equations. (\emph{see e.g.} \citet{Xue2005,Gu2009,Liu2010a}).

Similar to \citet{Narayan1995}, we adopt the self-similar assumptions in the radial direction to simplify the continuity and momentum equations, then reform and obtain the following vertical equations: \beq\ \frac{1}{2} {v_r}^2 + \frac{5}{2} {c_{\rm s}}^2 + {v_\phi}^2 - r^2 {\Omega_{\rm K}}^2 = 0, \eeq\ \beq\ \frac{1}{\rho} \frac{d p}{d \theta} = {v_\phi}^2 \cot \theta, \eeq\ \beq\ v_r = -\frac{3}{2} \frac{\alpha {c_{\rm s}}^2}{r \Omega_{\rm K}}, \eeq  where $v_r$ and $v_{\phi}$ are the radial and azimuthal components of the velocity ($v_\theta=0$), the sound speed $c_{\rm s}$ is defined as $c_{\rm s}^2 = p/\rho$, the Keplerian angular velocity is $\Omega_{\rm K} = (GM/r^3)^{1/2}$, and $\alpha$ is the constant viscosity parameter.

According the continuity equation, the mass accretion rate can be written as  \beq \dot{M} = -4 \pi r^2 \int_{\theta_0}^{\pi /2} \rho v_r \sin \theta d \theta, \eeq where $\theta_0$ is the polar angle of the surface and $\rho$ is the density.

The total pressure of gas is expressed as the summation of four pressures: \beq\ p = p_{\rm gas} + p_{\rm rad} + p_{\rm e} + p_\nu, \eeq where $p_{\rm gas}$, $p_{\rm rad}$, $p_{\rm e}$, and $p_\nu$ are the gas pressure from nucleons, the radiation pressure of photons, the degeneracy pressure of electrons, and the radiation pressure of neutrinos, respectively. Detailed expressions of the pressure components were given in \citet{Liu2007}. It should be pointed out that NDAFs are extremely optical thick for photons, so the radiation pressure of photons at anywhere can be written as \beq p_{\rm rad}= \frac{1}{3} a T^4, \eeq where $a$ is the radiation constant and $T$ is the temperature of gas. Additionally, we assume the polytropic relation in the vertical direction, $p = K \rho ^{4/3}$ , where $K$ is a constant and $\gamma=4/3$ is the adiabatic index.

Considering the energetic balance, the energy equation is written as \beq\ Q_{\rm vis} = Q_{\rm adv} + Q_\nu, \eeq where $Q_{\rm vis}$, $Q_{\rm adv}$, and $Q_\nu$ are the viscous heating, advection and neutrino cooling rates per unit area, respectively. In this paper, we ignore the cooling of photodisintegration of $\alpha$-particles and other heavier nuclei. The viscous heating rate per unit volume $q_{\rm vis} = \nu \rho r^2 [\partial (v_{\phi}/r)/\partial r]^2$ and the advective cooling rate per unit volume $q_{\rm adv} = \rho v_r (\partial e/\partial r - (p/\rho^2) \partial \rho/\partial r)$ ($e$ is the internal energy per unit volume) are expressed, after self-similarly simplification, as \beq\ q_{\rm vis} = \frac{9}{4} \frac{\alpha p v_{\phi}^2}{r^2 \Omega_{\rm K}}, \eeq \beq\ q_{\rm adv} = - \frac{3}{2} \frac{(p-p_{\rm e}) v_r}{r}, \eeq where the entropy of degenerate particles is neglected. Thus the vertical integration of $Q_{\rm vis}$ and $Q_{\rm adv}$ are the following: \beq\ Q_{\rm vis} = 2 \int_{\theta_0}^{\frac{\pi}{2}} q_{\rm vis}  r \sin {\theta} d\theta \ , \eeq \beq\ Q_{\rm adv} = 2 \int_{\theta_0}^{\frac{\pi}{2}} q_{\rm adv}  r \sin {\theta} d\theta . \eeq The cooling due to the neutrino radiation $Q_\nu$ can be defined as \citep{Lee2005} \beq\ Q_\nu = 2 \int_{\theta_0}^{\frac{\pi}{2}} q_\nu {\rm e}^{-\tau_\nu} r \sin {\theta} d\theta \ , \eeq where $q_\nu$ is the summation of the cooling rates per unit volume due to the Urca processes, electron-positron pair annihilation, nucleon-nucleon bremsstrahlung, and Plasmon decay (hereafter we represent then with $q_i$ ($i$=1, 2, 3, 4), respectively. Different type neutrinos involve different processes, for details, see \citet{Liu2007}); $\tau_\nu = \tau_{a, \nu} + \tau_{s, \nu}$ is the neutrino optical depth including absorption optical depth $\tau_{a, \nu}$ and scattering optical depth $\tau_{s, \nu}$, which can be defined as \beq \tau_{a, \nu} \approx \displaystyle{\sum_{i}} \frac{\int_{\theta_0}^{\theta} 2 q_i r d \theta}{7 \sigma T^4}, \eeq  \beq \tau_{s, \nu} \approx \displaystyle{\sum_{j}} \int_{\theta_0}^{\theta} \sigma_j n_j  r d \theta, \eeq where $\sigma_j$ and $n_j$ ($j=1$, 2, 3, 4) are the cross sections and the number density of protons, neutrons, $\alpha$-particles, and electrons, respectively (\emph{e.g.} \citet{Kohri2005,Chen2007}). In order to embody the complicated microphysics in NDAFs, we need to evaluate the electron fraction $Y_{\rm e} = n_{\rm p}/(n_{\rm p}+n_{\rm n})$ by solving equations (42), (44) and (45) in Liu et al. (2007).

\subsection{Boundary condition}

A boundary condition is required for solving the equations. The Eddington luminosity, which is defined by the equilibrium between the gravity and the radiation pressure, is closely related to the energy-liberation rate associated with mass-accretion processes. If the gradient of radiation pressure is larger than the gravity, outflows may occur, so we considered that it could present a natural boundary condition for the accretion disk. Deviated from \citet{Liu2010a}, here boundary condition is set to be the mechanical equilibrium, like solving principle of Eddington luminosity. Furthermore, we can ignore the radial gradient of radial velocity and pressure at the surface of the disk if the disk is thick. There are three forces to balance, namely gravity, radiation pressure force and centrifugal force in non-inertial reference frame. The mechanical equilibrium can be written as \beq p_{\rm rad} \mid _{\theta=\theta_0} \sigma_{\rm T} = \frac{2GMm_u}{r^2}\rm cot \theta_0, \eeq combined with eq. (6), the surface temperature can be defined as  \beq T_{\rm surf}=(\frac{6 G M m_{u}}{a \sigma_{\rm T} r^2} \rm cot \theta_0)^{\frac{1}{4}}, \eeq where $m_{\rm u}$ is the mean mass of a nucleon, $\sigma_{\rm T}$ is the Thompson scattering cross.

\begin{figure*}
\centering
\includegraphics[angle=0,scale=0.25]{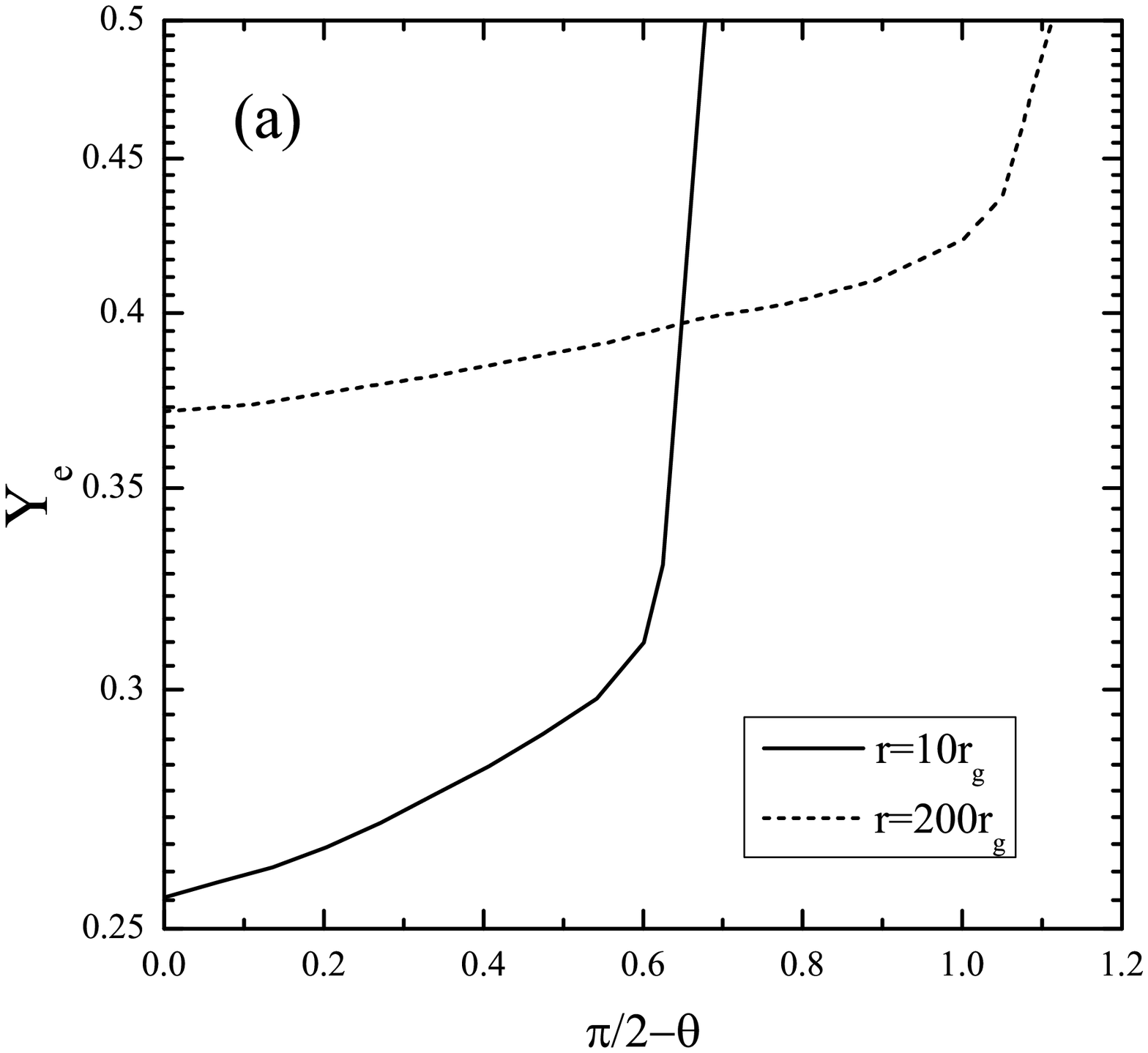}
\includegraphics[angle=0,scale=0.25]{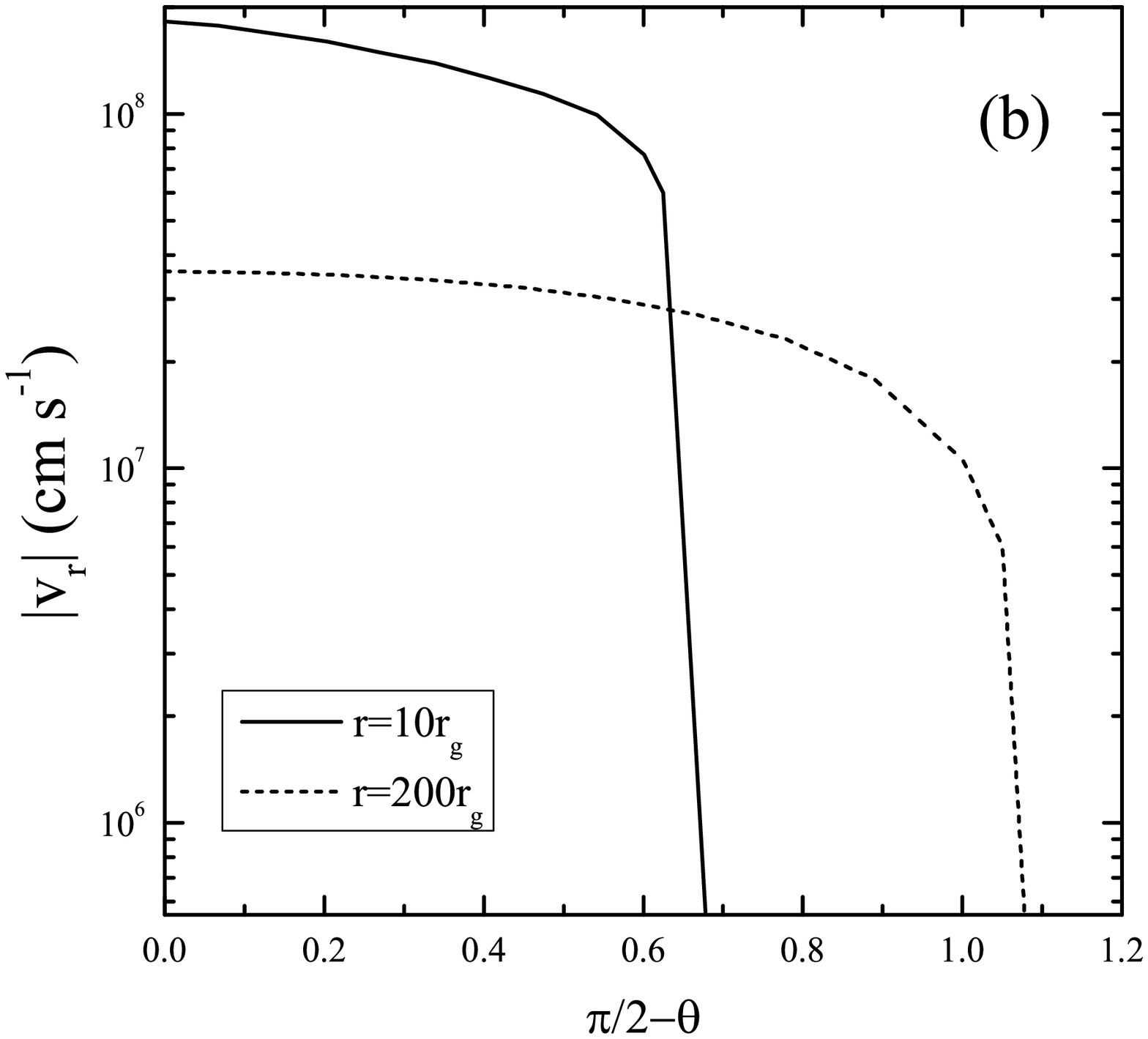}
\includegraphics[angle=0,scale=0.25]{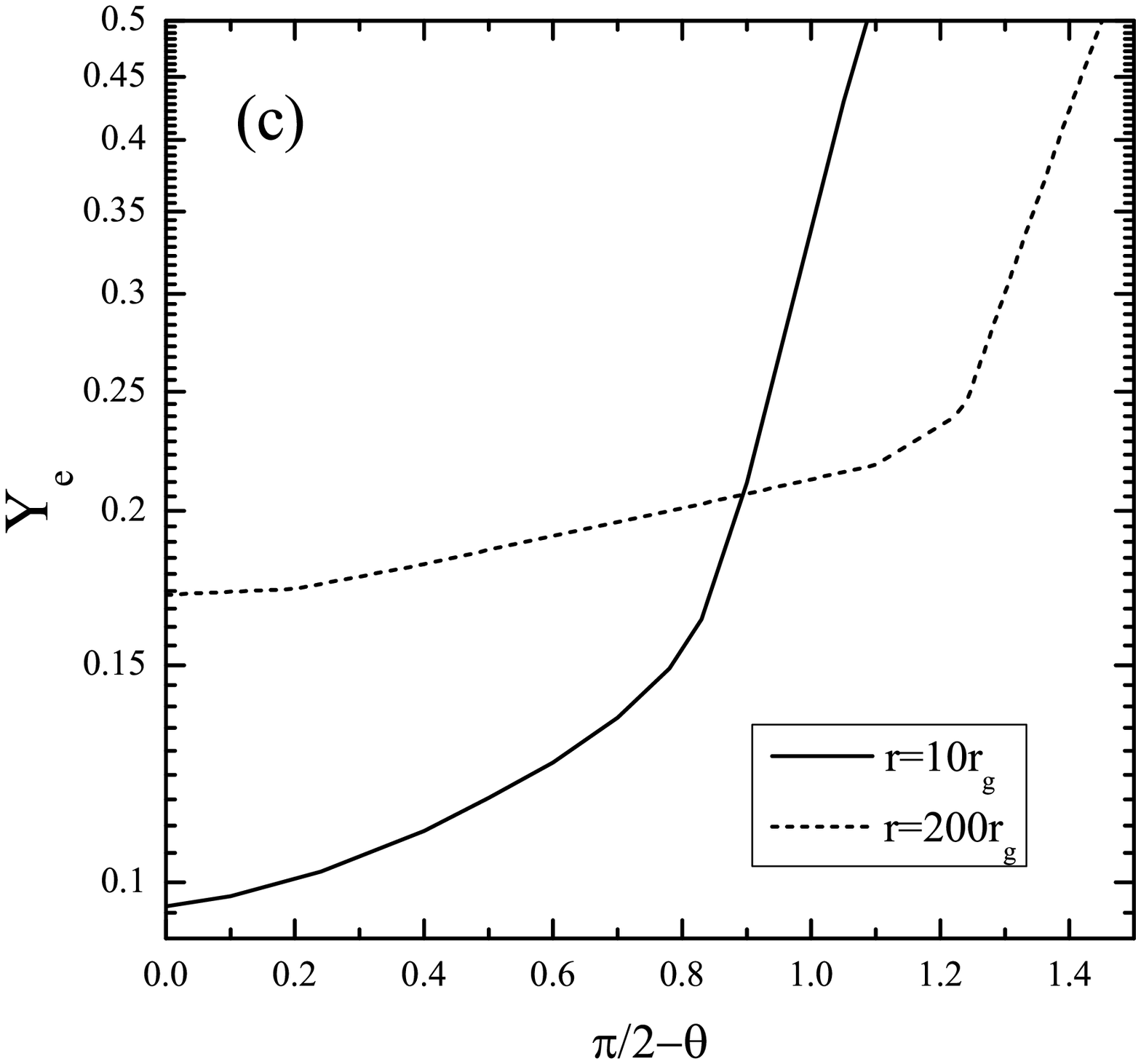}
\includegraphics[angle=0,scale=0.25]{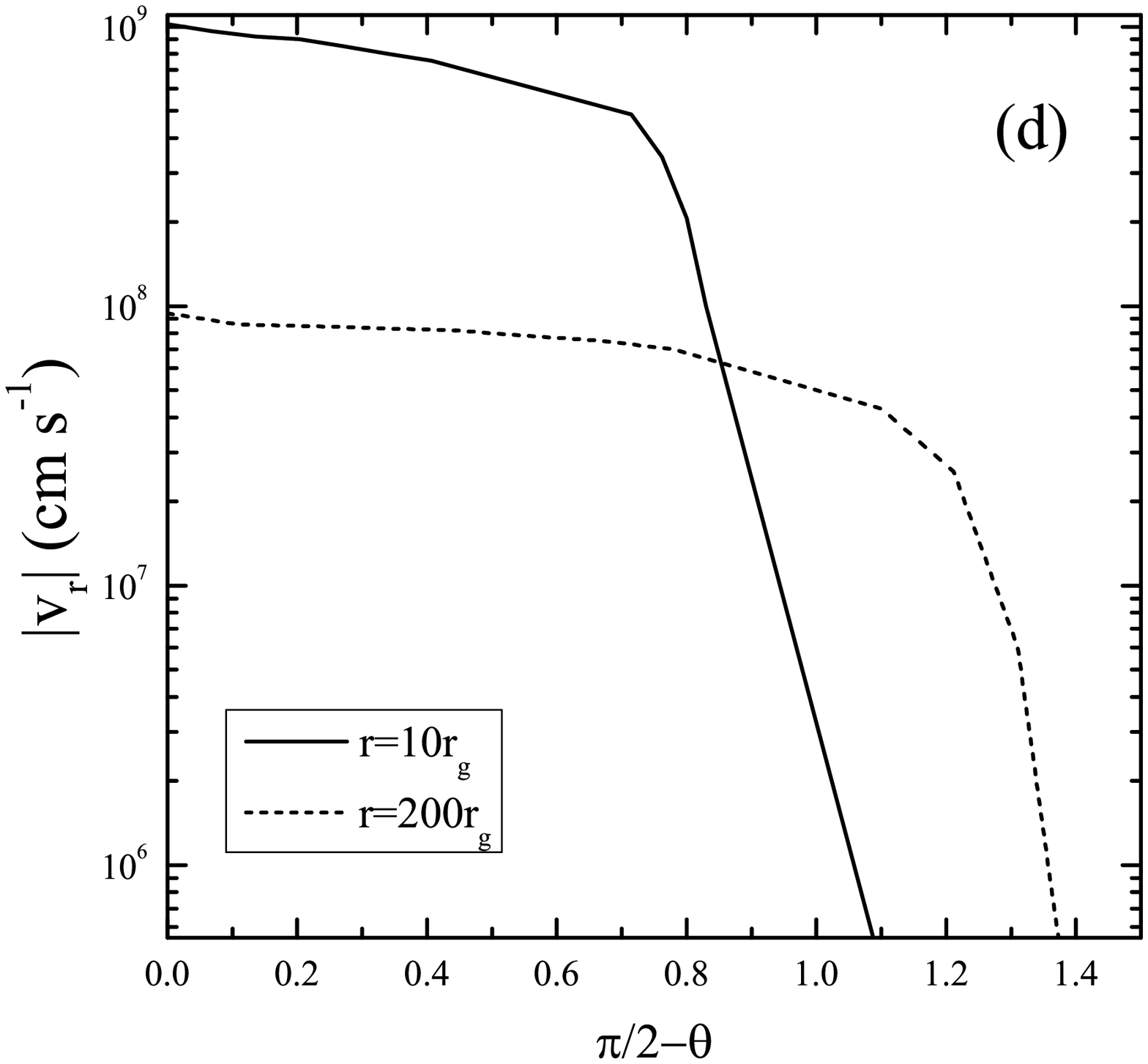}
\caption{Variations of the electron fraction $Y_{\rm e}$ and the radial velocity $v_r$ with the polar angle $\theta$ for which the given parameters are Case 1, comparing (a) and (b); Case 2, comparing (c) and (d).}
\label{sample-figure3}
\end{figure*}

\subsection{Bernoulli parameter and neutrino trapping}

The Bernoulli parameter of the accreted matter is expressed as (\emph{e.g.} \citet{Narayan1995}) \beq B=\frac{\gamma}{\gamma-1}{c_{\rm s}}^2 + \frac{1}{2}({v_r}^2+{v_\phi}^2)-\frac{GM}{r}, \eeq which includes the kinetic energy, potential energy and enthalpy of accreted gas.

In classic accretion theory, the radiation energy generated near the equatorial plane diffuses toward the disk surface at the speed of $\sim c/3 \tau$ \citep{Mihalas1984}, where $\tau$ is the total optical depth. Thus, the timescale of radiative diffusion is $t_{\rm diff}= H / (c/3 \tau)$ \citep{Ohsuga2002}, where $H$ is the half thickness of disk. We get $v_n$ for neutrinos replace $c$ for photons. Obviously, $v_n$ is related to the energy of neutrino $\sim 3.7 k T$ (\emph{e.g.} \citet{Di Matteo2002,Liu2007}). Since the accretion timescale $t_{\rm acc}$ is given by $-r/v_r$, the condition in which the neutrino radiation energy in the disk is trapped in the flow and falls onto black hole is written as $t_{\rm diff}> t_{\rm acc}$. Here we define a parameter as \beq \tilde{t} = \frac{t_{\rm diff}}{t_{\rm acc}} \approx - \frac{3 H \tau_\nu v_r}{r v_n}, \eeq only the optical depth of electron neutrino is calculated.

We numerically solve equations (1-12) with the boundary condition (16) for $\rho$, $v_r$, $v_{\phi}$, $c_s$, $p$, and $T$ by given $\alpha$, $M$, $\dot{M}$, and $r$(in this paper, we only concentrate on the stellar black hole with $M=3M_\odot$).

\section{Numerical Results and Possible Applications to GRBs}

Figure 1 shows the variations of the half-opening angles of the disk surface $(\pi/2-\theta_0)$ (solid lines), $B=0$ (dashed lines), $\tilde{t}=1$ (dotted line) with radii $r/r_g$ for which the given parameters are set as (Case 1) $\alpha=0.1$ and $\dot{M}=1~M_\odot ~\rm s^{-1}$, comparing (a); (Case 2) $\alpha=0.01$ and $\dot{M}=10~M_\odot ~\rm s^{-1}$, comparing (b). Here $r_g = 2GM/c^2$ is the Schwarzschild radius. In figure 1(a), the variation of half-opening angle from $\sim 0.56$ to $\sim 1.25$ with $r$ from $3r_g$ to $500r_g$. The region of $B>0$ appears at $r \sim 10r_g$, and expands continuously with the increase of radius. There is no region of neutrino trapping. In figure 1(b), the variation of half-opening angle from $\sim 0.91$ to $\sim 1.50$. The region of $B>0$ appears at $\sim 9r_g$, and increases to near the surface of the disk at $\sim 500r_g$. The region of neutrino trapping occurs from inner area to $\sim 46r_g$. We also notice the variation on opening angle of $\tilde{t}=1$ is from $\sim 0.4$ to $0$. We consider that the radius and the open angle of $B=0$ and $\tilde{t}=1$ is closely determined to the accretion rate and the viscous parameter \citep{Chen2007}.

The existence of the $B>0$ region implies the possible outflow. $B<0$, everywhere is a sufficient condition for the absence of outflows \citep{Abramowicz2000}. The neutrino trapping can decrease the neutrino radiation luminosity and annihilation luminosity. In the case of Figure 1(b), the neutrino luminosity is obtained as \beq L_\nu = 4 \pi \int_{3r_g}^{500r_g} Q_\nu r dr, \eeq which decreases $\sim 18\%$ in comparison with \citet{Liu2010a}. It is still about $10^{54} \rm ergs$ $\rm s^{-1}$ in Case 2 to produce GRBs according to neutrino annihilation. The disk is still thick, thus the volume above the disk shrinks and the radiated neutrino density increases. The neutrino annihilation luminosity is also larger than the results of the previous works (\citet{Popham1999,Di Matteo2002,Gu2006,Liu2007}). The 3-dimensional numerical simulations have shown that the neutrino annihilation luminosity is likely to be adequate for GRBs (\citet{Ruffert1999,Rosswog2003}). Our model is an acceptable estimation in comparison with the simulation results.

 Figure 2 shows the variations of the electron fraction $Y_{\rm e}$ and the radial velocity $v_r$ with the polar angle $\theta$ for which the given parameters are Case 1, comparing (a) and (b); Case 2, comparing (c) and (d). The profiles of $v_r$ are similar to that of the optically thin advection-dominated accretion flows \citep{Narayan1995}. Figures 2(a) and 2(c) show the variety of electron fraction in the vertical direction. Electron fraction $Y_{\rm e}$ of Case 1 near the equatorial plane is larger than Case 2, because it is determined by accretion rate more than viscosity parameter. $Y_{\rm e}$ is approach $0.5$ at the surface of the disk in the outer region, which means the matter of the disk is almost $\alpha$-particles and electrons in non-degenerate phase. GRBs may be the important contributors to local, and possibly galactic, nucleosynthesis, especially for rare species such as $r$-process nuclei. A lot of works have studied the nucleosynthesis in the central engine of GRBs (\emph{e.g.} \citet{Surman2004}). The distribution of electron fraction in our solutions may be an implication for the origin of nucleosynthesis in GRBs. In figures 2(b) and 2(d), the radial velocity $v_r$ is obviously less than it near the equatorial plane. It has not greatly changed between Case 1 and Case 2, because it is correlated positively with the accretion rate and viscosity parameter. The vertical structure of the disk is like a ``sandwich" (\emph{e.g.} \citet{Zhang2000}). The matter near the equatorial plane has been accreted first, inevitably, the remnants of the matter near the surface, which is almost in non-degenerate phase, has been accreted later, any radial disturbed motion will destroy the remnants to fracture in some parts. Two steps of accretion processes with different mass accretion rate correspond the GRB and its late flares. We find the similar model has been proposed by \citet{Lee2009}. They found that powerful winds are launched from the surface of the disk, driven by the recombination of free nucleons into $\alpha$-particles, which may produce the late-time X-ray flares.

\section{Conclusions and Discussion}

In this paper we revisit the vertical structure of NDAFs in spherical coordinates and calculate the vertical distributions of Bernoulli parameter, neutrino trapping, electron fraction and radial velocity. The major points we wish to stress are as follows:

\begin{enumerate}
\item We show the vertical structure of NDAFs under the mechanical equilibrium, and find that the flow will be significantly thick.

\item The Bernoulli parameter and neutrino trapping are determined by the mass accretion rate and the viscosity parameter.

\item According to the distribution of the Bernoulli parameter, we find that the possible outflow may appear, particularly in the outer region of the disk.

\item The neutrino trapping can essentially affect the neutrino radiation luminosity and the neutrino annihilation luminosity in turn.

\item The distributions of radial velocity and electron fraction indicate that the vertical structure of NDAF is like a ``sandwich", and this multilayer accretion may account for the progenitor of the late-time flares in GRBs.
\end{enumerate}

It is generally a two-dimensional problem to study a steady and axisymmetric accretion flow, and it is quite difficult for directly solving the two-dimensional system of partial differential equations. Most previous works (\emph{e.g.} \citet{Popham1999,Narayan2001,Kohri2002,Di Matteo2002,Kohri2005,Lee2005,Gu2006,Chen2007,Janiuk2007,Liu2007}) focused on the radial structure and therefore simplified the vertical counter-part of disk by using a one-zone approximation. The purpose of the present work is, however, to investigate the vertical structure in more details. Consequently, it requires us to directly solve the differential equations in the vertical direction, thus some simplification has to be taken in the radial direction. The self-similar assumption may be an excellent method to achieve the above purpose. Moreover, such an assumption has been widely adopted and successfully applied to neutrino-radiative disks (\emph{e.g.} \citet{Narayan2001}) as well as to photon-radiative disks (\emph{e.g.} \citet{Narayan1994}).

Based on the radial self-similar assumption, we can obtain some new results compared with the previous works. For example, \citet{Chen2007} discussed the radial distribution of Bernoulli parameter, neutrino trapping, and electron fraction, but the vertical distribution has to be ignored in that work. On the contrary, we can present the explicit vertical distribution of physical quantities as shown in the figures.

\citet{Xue2005} and \citet{Jiao2011} studied the outflows from accretion disks by assuming $v_\theta \neq 0$ in spherical coordinates. Their results indicated that the outflows
should be common in various accretion disks and might be stronger in slim disks, where both advection and radiation pressure are dominant. However, the radiation processes are not considered in these works. The new vertical structure of NDAFs should be calculated by reserving $v_\theta$ to discuss the natural outflow from the surface of the disk.

In our next work, we will take neutrino transfer effects (\emph{e.g.} \citet{Sawyer2003,Rossi2007}) into consideration instead of the polytropic equation of state, and calculate the variation of the neutrino luminosity and annihilation luminosity with the mass accretion rates. The ``sandwich" accretion disk model can also be studied, and some simulations should be done to discuss the possible origin of flares in GRBs. Moreover, we will study the complex nucleosynthesis process in the vertical direction of NDAFs.

\begin{acknowledgements}
We thank the anonymous referee for his/her valuable comments and constructive suggestions. This work was supported by the National Basic Research Program (973 Program) of China under grant 2009CB824800, the National Natural Science Foundation of China under grants 10833002, 11003016, 11073015 and 11103015, and the Natural Science Foundation of Fujian Province of China under grant 2010J01017.
\end{acknowledgements}

\end{document}